\documentclass[aps,preprintnumbers,floatfix,article,amsmath,amssymb,floatfix,10pt,prd,superscriptaddress,nofootinbib,scrartcl,twocolumn]{revtex4-2}
\usepackage{bm}
\usepackage{amsfonts}
\usepackage{latexsym}
\usepackage[latin1]{inputenc}
\usepackage{graphicx}
\usepackage{amsmath}
\usepackage{palatino}
\usepackage{mathpazo}
\usepackage[british]{babel}
\usepackage{hhline}
\usepackage{multirow}
\usepackage{textcomp}
\usepackage{float}
\usepackage{booktabs}
\usepackage{dcolumn}
\usepackage{hhline}
\usepackage{multirow}
\usepackage{ragged2e}
\usepackage{hyperref}
\hypersetup{colorlinks,citecolor=blue}
\hypersetup{colorlinks=true,linkcolor=red,filecolor=magenta,    urlcolor=cyan}
\usepackage{amsmath}
\usepackage{xcolor}
\usepackage{orcidlink}
\usepackage{epsfig}
\usepackage{caption}
\usepackage{subcaption}
\usepackage{commath}
\def\jnl@style{\it}
\def\aaref@jnl#1{{\jnl@style#1}}

\def\aaref@jnl#1{{\jnl@style#1}}

\def\aj{\aaref@jnl{AJ}}                   
\def\apj{\aaref@jnl{ApJ}}                 
\def\apjl{\aaref@jnl{ApJ}}                
\def\apjs{\aaref@jnl{ApJS}}               
\def\apss{\aaref@jnl{Ap\&SS}}             
\def\aap{\aaref@jnl{A\&A}}                
\def\aapr{\aaref@jnl{A\&A~Rev.}}          
\def\aaps{\aaref@jnl{A\&AS}}              
\def\mnras{\aaref@jnl{Mon.~Not.~Roy.~Astron.~Soc.}}             
\def\prd{\aaref@jnl{Phys.~Rev.~D}}        
\def\prc{\aaref@jnl{Phys.~Rev.~C}}  
\def\prl{\aaref@jnl{Phys.~Rev.~Lett.}}    
\def\qjras{\aaref@jnl{QJRAS}}             
\def\skytel{\aaref@jnl{S\&T}}             
\def\ssr{\aaref@jnl{Space~Sci.~Rev.}}     
\def\zap{\aaref@jnl{ZAp}}                 
\def\nat{\aaref@jnl{Nature}}              
\def\aplett{\aaref@jnl{Astrophys.~Lett.}} 
\def\apspr{\aaref@jnl{Astrophys.~Space~Phys.~Res.}} 
\def\physrep{\aaref@jnl{Phys.~Rep.}}      
\def\physscr{\aaref@jnl{Phys.~Scr}}       
\def\commat{\aaref@jnl{Comm.~Math.~Phys.}}              
\def\science{\aaref@jnl{Science}}               
\def\cqg{\aaref@jnl{Classical Quant.~Grav.}}            
\def\jpcs{\aaref@jnl{JPCS}}                                     
\def\ijmpd{\aaref@jnl{Int.~J.~Mod.~Phys.~D}}                    
\def\grg{\aaref@jnl{Gen.~Relat.~Gravit.}}               
\def\rpp{\aaref@jnl{Rep.~Prog.~Phys.}}          
\def\npa{\aaref@jnl{Nucl.~Phys.~A}}        
\def\lrr{\aaref@jnl{Living Rev.~Rel.}}                   
\def\jcap{\aaref@jnl{J.~Cosmology Astropart.~Phys.}}    
\def\rmp{\aaref@jnl{Rev.~Mod.~Phys.}}   
\def\epjc{\aaref@jnl{Eur.~Phys.~J.~C}} 
\def\plb{\aaref@jnl{~Phy.~Lett.~B}} 
\def\mpla{\aaref@jnl{Mod.~Phy.~Lett.~A}} 
\def\arxiv{\aaref@jnl{arxiv.org}}

\allowdisplaybreaks[1]

\begin{document}
\color{black}       
\title{\bf Exploring Late-Time Cosmic Acceleration: A Study of a Linear $f(T)$ Cosmological Model Using Observational Data}

\author{A. Zhadyranova} 
\email[Email: ]{a.a.zhadyranova@gmail.com}
\affiliation{General and Theoretical Physics Department, L. N. Gumilyov Eurasian National University, Astana 010008, Kazakhstan.}

\author{M. Koussour\orcidlink{0000-0002-4188-0572}}
\email[Email: ]{pr.mouhssine@gmail.com}
\affiliation{Department of Physics, University of Hassan II Casablanca, Morocco.} 

\author{S. Bekkhozhayev}
\email[Email: ]{s.o.bekkhozhayev@gmail.com}
\affiliation{General and Theoretical Physics Department, L. N. Gumilyov Eurasian National University, Astana 010008, Kazakhstan.}

\author{V. Zhumabekova}
\email[Email: ]{zh.venera@mail.ru}
\affiliation{Theoretical and Nuclear Physics Department, Al-Farabi Kazakh National University, Almaty, 050040, Kazakhstan.}

\author{J. Rayimbaev\orcidlink{0000-0001-9293-1838}}
\email[Email: ]{javlon@astrin.uz}
\affiliation{New Uzbekistan University, Mustaqillik Ave. 54, Tashkent 100007, Uzbekistan.}
\affiliation{University of Tashkent for Applied Sciences, Gavhar Str. 1, Tashkent 100149, Uzbekistan.}
\affiliation{National University of Uzbekistan, Tashkent 100174, Uzbekistan.}

\begin{abstract}
\textbf{Abstract}: Understanding the evolution of dark energy poses a significant challenge in modern cosmology, as it is responsible for the universe's accelerated expansion. In this study, we focus on a specific $f(T)$ cosmological model and analyze its behavior using observational data, including 31 data points from the CC dataset, 1048 points from the Pantheon SNe Ia samples, and 6 points from the BAO dataset. By considering a linear $f(T)$ model with an additional constant term, we derive the expression for the Hubble parameter as a function of cosmic redshift for non-relativistic pressureless matter. We obtain the best-fit values for the Hubble constant, $H_0$, and the model parameters $\alpha$ and $\beta$, indicating a stable model capable of explaining late-time cosmic acceleration without invoking a dark energy component. This is achieved through modifying field equations to account for the observed accelerated expansion of the universe.
 
\end{abstract}

\maketitle

\section{Introduction} \label{sec1}

During the previous 20 years, evidence from various cosmological observations has strongly supported the idea of an accelerating cosmological expansion. Type Ia Supernovae (SNe Ia) searches, such as those conducted by Riess et al. and Perlmutter et al., have given important new information on the universe's expansion history \cite{Riess,Perlmutter}. In addition, observations from the Wilkinson Microwave Anisotropy Probe (WMAP) experiment \cite{C.L.,D.N.} and measurements of the Cosmic Microwave Background Radiation (CMBR) \cite{R.R.,Z.Y.} have further corroborated these findings. Studies of Large-Scale Structure (LSS) in the universe, including galaxy surveys, have also contributed significantly to our understanding, with research by Koivisto et al. and others playing a key role \cite{T.Koivisto,S.F.}. Furthermore, measurements of Baryonic Acoustic Oscillations (BAO) \cite{D.J.,W.J.} have provided additional confirmation of the accelerating expansion, collectively painting a compelling picture of the dynamic nature of our universe. The ultimate destiny of the universe is a subject of significant interest, hinging on the presence of dark energy (DE), a form of energy with negative pressure believed to drive the current cosmological expansion. The future evolution of our universe is intricately linked to the fundamental properties of DE. One way to characterize DE is through its equation of state (EoS) parameter, denoted as $\omega_{DE}$, which is defined as the ratio of the spatially homogeneous pressure $p_{DE}$ to the energy density $\rho_{DE}$ of DE. Recent cosmological observations have not provided sufficient precision to distinguish between different scenarios for the EoS parameter of DE, where $\omega_{DE} < -1$, $\omega_{DE} = -1$, and $\omega_{DE} > -1$. For instance, the value of $\omega$ derived from the combination of data from the WMAP, including measurements of $H_0$, SNe Ia, CMB, and BAO, yields $\omega_{DE}=-1.084\pm0.063$ \cite{G.H.}. In 2015, the Planck collaboration reported $\omega_{DE}=-1.006\pm0.0451$ \cite{Planck/2015}, and in 2018, they revised this value to $\omega_{DE}=-1.028\pm0.032$ \cite{Planck/2018}.

The simplest explanation for DE in general relativity (GR) is the cosmological constant $\Lambda$, characterized by an EoS parameter $\omega_{\Lambda}=-1$. However, a significant disparity exists between the observed value of the cosmological constant $\Lambda$ and its expected value from quantum gravity \cite{S.W.}. This inconsistency is known as the cosmological constant problem. Another extensively studied time-varying DE model is the quintessence, characterized by an EoS $-1 < \omega_{DE} < -\frac{1}{3}$ \cite{RP,M.T.}. In quintessence models, the DE field evolves and can exhibit behavior ranging from nearly pressureless matter-like ($\omega \approx 0$) to that of a cosmological constant ($\omega_{\Lambda} = -1$). Further, DE characterized by $\omega_{DE}<-1$, known as phantom energy, represents one of the least theoretically understood forms of energy. Phantom energy has intriguing properties, such as negative kinetic energy and the potential to cause a big rip singularity in the future universe, where the expansion rate becomes infinite \cite{KI,DB,DB-2}. Despite its exotic nature, phantom energy is a valid theoretical possibility within the framework of GR and modified gravity theories.

In this study, we investigate an alternative approach where DE's evolution is driven by modifications to the gravitational sector rather than the matter sector. This approach is a cornerstone of modified theories of gravity, which have been extensively studied in the literature \cite{L.A.,SA,R.F.,Capozziello11}. These theories are typically constructed by extending the Einstein-Hilbert action, which is based on curvature. However, an alternative approach involves modifying the action of the equivalent torsional formulation of GR, such as the teleparallel equivalent of GR (TEGR) \cite{Einstein28,Maluf94,Unzicker05,Aldrovandi13}. In GR, the Levi-Civita connection is associated with curvature, while torsion is not considered. On the other hand, in teleparallelism, the Weitzenb$\ddot{o}$ck connection is associated with torsion, and curvature is not taken into account \cite{Weitzenbock23}. In this context, the fundamental entities are the four linearly independent tetrad fields, which serve as the orthogonal bases for the tangent space at each point in spacetime. The torsion tensor is constructed from the products of the first derivatives of the tetrad. The $f(R)$ gravity is a basic modification of GR that maintains zero torsion, similar to $f(T)$ gravity, which is a basic modification of the TEGR with zero curvature. It's widely recognized that the action of general teleparallel gravity doesn't adhere to local Lorentz invariance. \cite{Sotiriou11}. In $f(T)$ gravity formulation \cite{Bengochea09, Ferraro08, Linder10}, one starts with pure tetrad teleparallel gravity, where the spin connection is assumed to be zero. This results in the torsion tensor being effectively replaced by coefficients of the governing parameters, which are not tensors under local Lorentz transformations. While the violation of local Lorentz symmetry is overlooked in TEGR since it doesn't impact the field equations, it becomes a concern in $f(T)$ gravity. It's noteworthy that teleparallel gravity employs the teleparallel connection $\Gamma^{\rho}_{~~\mu \nu}$, including torsion but no curvature \cite{Krssak19,Bahamonde21}. In contrast, curvature-based geometries use the Levi-Civita connection $\tilde{\Gamma}^{\rho}_{~~\mu \nu}$, which features non-vanishing curvature. Both connections are metric-compatible. Several studies in $f(T)$ gravity have explored various aspects, including cosmological solutions \cite{Paliathanasis/2016}, thermodynamics \cite{Salako/2013}, late-time acceleration \cite{Myrzakulov/2011, Bamba/2011}, cosmological perturbations \cite{Chen/2011}, large-scale structure \cite{Li/2011}, cosmography \cite{Capozziello/2011}, energy conditions \cite{Liu/2012}, matter bounce cosmology \cite{Cai/2011}, wormholes \cite{Jamil/2013}, anisotropic universe \cite{Rodrigues/2016,Koussour/2022}, and observational constraints \cite{Nunes/2016}.

Recently, there has been a significant increase in observational data, particularly from SNe Ia, BAO, and CMB. These datasets have given important new information about the universe's expansion. The Cosmic Chronometer (CC) dataset, in particular, has revealed the intricate structure of cosmic expansion. Measurements of the ages of the most massive galaxies have directly measured the Hubble parameter $H(z)$ at various redshifts $z$, introducing a new standard cosmological probe. In this study, we incorporate 31 measurements of CC obtained using the differential age method \cite{cc}, as well as the Pantheon dataset by Scolnic et al. \cite{Scolnic/2018}, which includes 1048 SNe Ia measurements spanning a redshift range of $0.01 < z < 2.3$. In addition, we use the BAO data comprising six points \cite{BAO1}. 

This paper is structured as follows. In Sec. \ref{sec2}, we introduce the mathematical formalism of $f(T)$ teleparallel gravity and its Friedmann-Lema\^{i}tre-Robertson-Walker (FLRW) cosmology. In Sec. \ref{sec3}, we consider a linear $f(T)$ model and derive expressions for the Hubble parameter, energy density, EoS parameter, and deceleration parameter. Sec. \ref{sec4} focuses on constraining the model parameters using 31 points from the CC dataset, 1048 points from the Pantheon SNe Ia samples, and 6 points from the BAO dataset. In addition, we examine the physical behavior of cosmological parameters. In Sec. \ref{sec5}, we analyze the stability of the cosmological $f(T)$ model using the scalar perturbation approach. Finally, in Sec. \ref{sec6}, we present our conclusions.

\section{Mathematical formalism of $f(T)$ teleparallel gravity} \label{sec2}

In teleparallel gravity, GR can be redefined by employing the tetrads as the dynamic variable instead of the metric tensor \cite{Einstein28}. The tetrad consists of a basis $\lbrace \textbf{e}_A({\textbf{x})\rbrace}$, where $A=0,1,2,3$, of vectors in spacetime. Every vector $e_A$  can be expressed in terms of a coordinate basis, yielding its components $e_a^{\mu}$. Therefore, the condition of orthogonality is transformed into,
\begin{equation}\label{1}
g_{\mu \nu}=\eta_{AB} e_{\mu}^{A} e_{\nu}^{B}.
\end{equation}

Here, $g_{\mu \nu}$ represents the metric tensor, and $\eta_{AB}=diag(1,-1,-1,-1)$ is the Minkowski metric tensor. The dual frame $\lbrace \textbf{e}^A\rbrace$ allows us to express the inverse of Eq. \eqref{1} as, $e^{\mu}_Ae^B_{\mu}=\delta_A^B$. Eq. \eqref{1} allows us to express $e=det[e^A_{\mu}]=\sqrt{-g}$. In $f(T)$ gravity \cite{Weitzenbock23}, the connection can be expressed as, $ \Gamma^{\lambda}_{\nu \mu}\equiv e^{\lambda}_{A} \partial_{\mu} e^{A}_{\nu}$ where the torsion tensor is represented by,
\begin{equation}\label{2}
T^{\lambda}_{\mu \nu}\equiv\hat\Gamma^{\lambda}_{\nu \mu}-\hat\Gamma^{\lambda}_{\mu \nu}=e^{\lambda}_{A} \partial_{\mu} e^{A}_{\nu}-e^{\lambda}_{A} \partial_{\nu} e^{A}_{\mu}.
\end{equation}

The torsion scalar is derived by contracting the torsion tensor,
\begin{equation}\label{3}
T \equiv \frac{1}{4} T^{\rho \mu \nu} T_{\rho \mu \nu}+\frac{1}{2} T^{\rho \mu \nu} T_{\nu \mu \rho}-T_{\rho \mu}^{~~\rho} T^{\nu \mu}_{~~\nu}.
\end{equation}

Further, the action for teleparallel gravity can be formulated from the teleparallel Lagrangian. In $f(T)$ gravity, $T$ is generalized to a function $T+f(T)$, and its action can be expressed as \cite{Anagnostopoulos19},  
\begin{equation}\label{4}
S = \frac{1}{16 \pi G}\int d^{4}xe[T+f(T)+\mathcal{L}_{m}],
\end{equation}
where $f(T)$ is an arbitrary function of the torsion scalar $T$, $G$ is the gravitational constant, and for completeness, we have included the matter Lagrangian $\mathcal{L}_{m}$. We adopt the natural system of units where $G=c =1$. 

By varying the action \eqref{4} with respect to the vierbein, we can derive the gravitational field equations as
\begin{equation}\label{5_split}
\begin{aligned}
& e^{-1}\partial_{\mu}(e e^{\rho}_{A}S_{\rho}^{~\mu \nu})[1+f_{T}] \\
& \quad + e^{\rho}_{A}S_{\rho}^{~\mu \nu}\partial_{\mu}(T)f_{TT} - e^{\lambda}_{A}T^{\rho}_{~\mu \lambda}S_{\rho}^{~ \nu\mu}[1+f_{T}] \\
& \quad + \frac{1}{4}e^{\nu}_{A}[T+f(T)] = 4 \pi e^{\rho}_{A}\mathcal{T}_{~\rho}^{~~\nu}.
\end{aligned}
\end{equation}

From now on, we will denote $f=f(T)$ with $f_{T}$ and $f_{TT}$ representing the first and second-order derivatives with respect to $T$ respectively. In addition, $\mathcal{T}_{~\rho}^{~~\nu}$ represents the energy-momentum tensor for the cosmic matter content. Moreover, the superpotential is defined as, 
\begin{equation}
S_{\rho}^{~~\mu \nu}\equiv\frac{1}{2}(K^{\mu \nu}_{~~~\rho}+\delta^{\mu}_{\rho}T^{\alpha \nu}_{~~~\alpha}-\delta^{\nu}_{\rho}T^{\alpha \mu}_{~~~\alpha}),    
\end{equation}
where the contortion tensor is given by \cite{Hehl76},
\begin{equation}
K^{\mu \nu}_{~~~\rho}\equiv \frac{1}{2}(T^{\nu \mu}_{~~~\rho}+T_{\rho}^{~~\mu \nu}-T^{\mu \nu}_{~~~\rho}).    
\end{equation}

The FLRW metric describes a homogeneous and isotropic universe on large scales, as predicted by the cosmological principle. It is a key element in cosmology for modeling the expansion of the universe. The metric is given by \cite{Ryden}
\begin{equation}
ds^2 = dt^2 - a(t)^2 \left[ \frac{dr^2}{1-kr^2} + r^2(d\theta^2 + \sin^2\theta d\phi^2) \right] \label{6}
\end{equation}
where $a(t)$ (denoted as the scale factor) is a function of cosmic time $t$ that describes how the distances between galaxies change as the universe expands, $k$ is the curvature parameter (which can be 1, 0, or -1 corresponding to closed, flat, or open universes respectively), and $e^{A}_{\mu}\equiv diag(1,a(t),a(t),a(t))$. Moreover, we can find the torsion scalar for the line element \eqref{6} as $T=- 6H^2$. In this context, $H = \dot{a}/a$ represents the Hubble parameter, indicating the rate at which the universe is expanding.

Furthermore, we assume the universe is filled with a perfect fluid-type matter, a theoretical construct in cosmology used to model matter distribution. A perfect fluid is defined by its lack of viscosity and zero thermal conductivity. Its energy-momentum tensor is described by
\begin{equation}\label{2c}
T_{\mu\nu}=(\rho+p)u_\mu u_\nu - pg_{\mu\nu},
\end{equation}
where $\rho$ denotes the matter-energy density, $p$ is the isotropic pressure, $u^\mu$ represents the four-velocity components with $u^\mu u_\mu=1$, and the trace of the energy-momentum tensor is given by  $T=\rho-3p$.

The modified Friedmann equations are fundamental for understanding the universe's dynamics within $f(T)$ gravity. These equations dictate how the scale factor $a(t)$ evolves. In the framework of $f(T)$ gravity, the modified Friedmann equations in a flat ($k=0$) FLRW spacetime can be expressed as
\begin{eqnarray}
16\pi \rho&=&f-T-2Tf_T, \label{F1}\\
\dot{H}&=&-\frac{4\pi (\rho+p)}{1+f_T+2Tf_{TT}}. \label{F2} 
\end{eqnarray}

Specifically, when considering the function $f(T)=0$, we can retrieve the standard Friedmann equations of GR. Now, taking the trace of the field equations yields the following equation for matter conservation,
\begin{equation}\label{3f}
\dot{\rho} + 3H\left(\rho+p\right)=0
\end{equation}

These equations \eqref{F1} and \eqref{F2} can be understood as the TEGR cosmology, with an extra component arising from the spacetime torsion that mimics the behavior of a DE fluid. These DE components arising from torsion are characterized by
\begin{eqnarray}
\rho_{DE}&\equiv&\frac{1}{16 \pi}\left[-f+2Tf_{T}\right], \label{9} \\
p_{DE}&\equiv&-\frac{1}{16 \pi}\left[\frac{-f+Tf_{T}-2T^{2}f_{TT}}{1+f_{T}+2Tf_{TT}}\right].\label{10} 
\end{eqnarray}

Further, the equation of state (EoS) parameter, which connects the energy density and pressure of the DE component, can be expressed as,
\begin{equation}\label{11}
\omega_{DE}=\frac{p_{DE}}{\rho_{DE}}=-1+\frac{\left(f_T+2Tf_{TT}\right)\left(-f+T+2Tf_T\right)}{(1+f_{T}+2Tf_{TT})(-f+2Tf_{T})}.
\end{equation}

Therefore, the effective Friedmann equations, including a component representing DE due to torsion, can be described as
\begin{equation}\label{3n}
H^2=\frac{8\pi}{3} \left[ \rho+\rho_{DE} \right]
\end{equation}
\begin{equation}\label{3o}
2\dot{H}+3H^2=-\frac{8\pi}{3} \left[p+p_{DE} \right]
\end{equation}

\section{Cosmological $f(T)$ model}\label{sec3}

To account for the cosmic acceleration in a homogeneous and isotropic universe, the late universe must exhibit significant negative pressure. Various models of DE have been proposed in the literature to explain this phenomenon. While the cosmological constant, $\omega_{\Lambda} = -1$, remains the most successful model in explaining the universe's accelerating expansion, it faces challenges such as the cosmological constant problem and the fine-tuning required for the observed value of $\Lambda$. To circumvent these issues, a time-dependent EoS parameter has been suggested. In this paper, we address this by considering the following linear $f(T)$ model, with the addition of a constant term:
\begin{equation}
    f(T)=\alpha T+\beta,
\end{equation}
where $\alpha$ and $\beta$ are parameters of the model. Recently, Maurya \cite{Maurya/2022} studied scenarios of cosmic acceleration in a universe filled with viscous fluid, employing the framework of modified $f(T)$ gravity. The investigation focused on using a linear function of the torsion scalar within this modified gravity theory. In addition, Shekh and Chirde \cite{Shekh/202O} presented an accelerating Bianchi-type DE cosmological model incorporating a cosmic string within the framework of linear $f(T)$ gravity. In the action (\ref{3}), $f(T)$ gravity is represented by $T + f(T)$. Thus, the function can be expressed as $T + \alpha T + \beta = (\alpha + 1)T + \beta$. Then, we can immediately deduce that $f_T=\alpha$ and $f_{TT}=0$. Therefore, for this particular $f(T)$ model, Eqs. (\ref{9})-(\ref{11}) simplify to
\begin{eqnarray}
\rho_{DE}&=&-\frac{ 6 \alpha  H^2 +\beta}{16 \pi },\label{14}\\
p_{DE}&=&\frac{\beta }{16 \pi  (\alpha +1) }, \label{15}\\
\omega_{DE} &=&-\frac{\beta }{(\alpha +1) \left(6 \alpha  H^2+\beta\right)}.\label{16}
\end{eqnarray}

Therefore, the effective EoS parameter $\omega_{eff}= \frac{p_{eff}}{\rho_{eff}}$ is given by
\begin{equation}\label{4f}
\omega_{eff}=\frac{p_{DE}}{\rho + \rho_{DE}}=\frac{\beta }{6 (\alpha +1) H^2}.
\end{equation}

To study the evolution of the cosmological parameters mentioned above, we must obtain an expression for $H$ for our model. Using Eq. (\ref{F2}), we obtain a first-order differential equation for a universe dominated by non-relativistic pressureless matter ($p=0$), expressed as
\begin{equation} \label{20}
\dot{H}+\frac{3}{2}H^2+\frac{\beta }{4 (\alpha +1)}=0.
\end{equation}

Using the relationship $ \frac{1}{H} \frac{d}{dt}= \frac{d}{dln(a)}$ (where $a(t)=\frac{1}{1+z}$) and then integrating Eq. (\ref{20}), we derive the expression for the Hubble parameter in terms of redshift $z$ as follows:
\begin{equation}\label{Hz}
H(z)=\sqrt{\frac{\left( 6H_{0}^2 (\alpha +1)+\beta\right)}{6(\alpha +1)}(1+z)^3-\frac{\beta}{6(\alpha +1)} },
\end{equation}
where $H(z=0) = H_0$ represents the present value of the Hubble parameter. In the scenario where $\alpha = 0$ and $\beta = 0$, i.e., $f(T)=0$, the solution simplifies to $H(z) = H_0 (1 + z)^{\frac{3}{2}}$, which describes a matter-dominated universe.

To discern if the cosmological expansion accelerates or decelerates, we introduce the deceleration parameter $q$, defined as
\begin{equation} \label{q}
q=-1-\frac{\dot{H}}{H^2}= -1+\frac{(1+z)}{H(z)}\frac{dH(z)}{dz}.
\end{equation}

Using Eqs. (\ref{Hz}) and (\ref{q}), we get
\begin{equation} \label{qz}
q(z)=-1+\frac{3 \left[6H_{0}^2 (\alpha +1)+\beta\right](1+z)^3}{2 \left[(1+z)^3 \left(6H_{0}^2 (\alpha +1)+\beta\right)-\beta \right]}.    
\end{equation}

From Eq. (\ref{Hz}), it is clear that the form of $H(z)$ obtained involves several model parameters and therefore critically relies on the values of these parameters. One could, in principle, select the model parameters arbitrarily and examine the functional behavior of the cosmological parameters derived. These arbitrarily chosen values could then be compared with observational data. However, we take a different approach. We first constrain the various model parameters using the available datasets. With the best-fit values obtained, we then investigate the behavior of $H(z)$ and $q(z)$.

\section{Data, Methodology, and Cosmological parameters}\label{sec4}

\subsection{Data and methodology}

To determine the model parameters $H_0$, $\alpha$, and $\beta$ in our cosmological model, we rely on the latest data from the CC, SNe Ia, and BAO observations. We employ 31 data points from the CC dataset, 1048 points from the Pantheon SN samples, and 6 points from the BAO dataset. We use Bayesian analysis and the likelihood function in conjunction with the Markov Chain Monte Carlo (MCMC) simulations, implemented using the \texttt{emcee} Python library \cite{Mackey/2013}.

\text{\textbf{Cosmic Chronometer dataset}}: The Hubble parameter, expressed as $H(z)=-dz/[dt(1+z)]$, allows for a model-independent determination of its value from observational data. Since $dz$ is obtained from a spectroscopic survey, measuring $dt$ enables the calculation of the Hubble parameter without reliance on a specific cosmological model. In this study, we use a collection of 31 $H(z)$ data points obtained from the differential age technique to avoid covariance matrix issues and potential correlations with the BAO data. The complete list of these 31 data points can be found in \cite{cc}.

\text{\textbf{SNe Ia dataset}}: Originally, observational studies on a sample of 50 SNe Ia points indicated that our universe is undergoing acceleration. Over the past two decades, studies on larger samples of Type Ia SNe datasets have increased. In this study, we analyze a sample of 1048 spectroscopically confirmed SNe Ia, referred to as the Pantheon dataset. Recently, Scolnic et al. \cite{Scolnic/2018} compiled the Pantheon dataset, which includes 1048 SNe Ia in the redshift range $0.01 < z < 2.3$. Contributions to the Pantheon dataset come from the PanSTARRS1 Medium Deep Survey, SDSS, SNLS, as well as numerous low-z and HST samples.

\text{\textbf{BAO dataset}}: The BAO dataset includes measurements from the 6dFGS, SDSS, and WiggleZ surveys, providing BAO measurements at six different redshifts, as shown in Tab. \ref{tab}. The characteristic scale of BAO is determined by the sound horizon $r_{s}$ at the photon decoupling epoch $z_{\ast }$, which is related by the following equation:
\begin{equation}\label{5f}
r_{s}(z_{\ast })=\frac{c}{\sqrt{3}}\int_{0}^{\frac{1}{1+z_{\ast }}}\frac{da}{
a^{2}H(a)\sqrt{1+(3\Omega _{0b}/4\Omega _{0\gamma })a}},
\end{equation}
where $\Omega _{0b}$ and $\Omega _{0\gamma }$ indicate the present densities of baryons and photons, respectively. In this study, the BAO datasets, consisting of six points for $d_{A}(z_{\ast })/D_{V}(z_{BAO})$ are sourced from Refs. \cite{BAO1, BAO2, BAO3, BAO4, BAO5, BAO6}. Here, we consider the redshift at the photon decoupling epoch as $z_{\ast }\approx 1091$. The co-moving angular diameter distance is defined as $d_{A}(z)=\int_{0}^{z}\frac{dz^{\prime }}{H(z^{\prime })} $, and the dilation scale $D_{V}(z)=\left(
d_{A}(z)^{2}z/H(z)\right) ^{1/3}$.

In our MCMC study, we use 100 walkers and 1000 steps to determine the fitting outcomes. We incorporate the following priors: $H_0 \in [60,80] , \text{km/s/Mpc}, \quad \alpha \in [-2,2], \quad \zeta_{0} \in [-30,30], \quad \zeta_{1} \in [-2,2]$. Moreover, for the combined $CC+SN+BAO$ dataset, we calculate the Chi-square and likelihood ($\mathcal{L} \propto exp(-\chi^2/2)$) as follows:
\begin{eqnarray}
\chi^{2}_{joint} &=& \chi^{2}_{CC} + \chi^{2}_{SN}+ \chi^{2}_{BAO},\\
\mathcal{L}_{joint} &=& \mathcal{L}_{CC} \times \mathcal{L}_{SN} \times \mathcal{L}_{BAO},
\end{eqnarray}
where
\begin{eqnarray}
\chi^{2}_{CC} &=& \sum_{i=1}^{31} \frac{\left[H(\theta_{s}, z_{i})-
H_{obs}(z_{i})\right]^2}{\sigma(z_{i})^2},\\
\chi^{2}_{SN} &=& \sum_{i,j=1} ^{1048} \Delta \mu_{i} \left(
C_{SN}^{-1}\right)_{ij} \Delta \mu_{j},\\
\chi _{BAO}^{2} &=& X^{T}C_{BAO}^{-1}X\,.
\end{eqnarray}

For $\chi^{2}_{H(z)}$, $H(\theta_{s}, z_{i})$ denotes the theoretical value of the Hubble parameter at redshift $z_{i}$, determined by the model parameters $\theta_{s}=(H_0, \alpha, \beta)$, $H_{obs}(z_{i})$ denotes its observed value at redshift $z_{i}$, and $\sigma(z_{i})$ is the standard error associated with the observed value of $H$. 

For $\chi^{2}_{SN}$, $\Delta \mu_{i}=\mu^{\rm th}(z_{i},\theta_{s})-\mu_i^{obs}$ represents the difference between the distance modulus of the $i_{th}$ SNe data point and its corresponding theoretical prediction, while $C_{SN}^{-1}$ denotes the inverse covariance matrix of the Pantheon sample. In addition,we define the calculated theoretical value of the distance modulus as: $\mu(z)=5log_{10}\frac{d_{L}(z)}{1Mpc}+25$, where $d_{L}(z)=c(1+z)\int_{0}^{z}\frac{dy}{H(y,\theta_{s} )}$ is the luminosity distance for a flat universe \cite{Planck/2018}. Here, $c$ is the speed of light.

For $\chi^{2}_{BAO}$, the vector $X$ varies depending on the specific survey under consideration,
\begin{equation*}
X=\left( 
\begin{array}{c}
\frac{d_{A}(z_{\star })}{D_{V}(0.106)}-30.95 \\ 
\frac{d_{A}(z_{\star })}{D_{V}(0.2)}-17.55 \\ 
\frac{d_{A}(z_{\star })}{D_{V}(0.35)}-10.11 \\ 
\frac{d_{A}(z_{\star })}{D_{V}(0.44)}-8.44 \\ 
\frac{d_{A}(z_{\star })}{D_{V}(0.6)}-6.69 \\ 
\frac{d_{A}(z_{\star })}{D_{V}(0.73)}-5.45%
\end{array}%
\right) \,,
\end{equation*}
while $C_{BAO}^{-1}$ represents the inverse covariance matrix for the BAO dataset \cite{BAO6},

\begin{widetext}

\begin{equation*}
C_{BAO}^{-1}=\left( 
\begin{array}{cccccc}
0.48435 & -0.101383 & -0.164945 & -0.0305703 & -0.097874 & -0.106738 \\ 
-0.101383 & 3.2882 & -2.45497 & -0.0787898 & -0.252254 & -0.2751 \\ 
-0.164945 & -2.454987 & 9.55916 & -0.128187 & -0.410404 & -0.447574 \\ 
-0.0305703 & -0.0787898 & -0.128187 & 2.78728 & -2.75632 & 1.16437 \\ 
-0.097874 & -0.252254 & -0.410404 & -2.75632 & 14.9245 & -7.32441 \\ 
-0.106738 & -0.2751 & -0.447574 & 1.16437 & -7.32441 & 14.5022%
\end{array}%
\right) \,.
\end{equation*}

\begin{table}[H]
\begin{center}
\begin{tabular}{|c|c|c|c|c|c|c|}
\hline
$z_{BAO}$ & $0.106$ & $0.2$ & $0.35$ & $0.44$ & $0.6$ & $0.73$ \\ \hline
$\frac{d_{A}(z_{\ast })}{D_{V}(z_{BAO})}$ & $30.95\pm 1.46$ & $17.55\pm 0.60$
& $10.11\pm 0.37$ & $8.44\pm 0.67$ & $6.69\pm 0.33$ & $5.45\pm 0.31$ \\ 
\hline
\end{tabular}
\caption{List of six values representing the ratio $d_{A}(z_{\ast })/D_{V}(z_{BAO})$ from BAO measurements.}
\label{tab}
\end{center}
\end{table}

\begin{figure}[H]
\centering\includegraphics[scale=0.98]{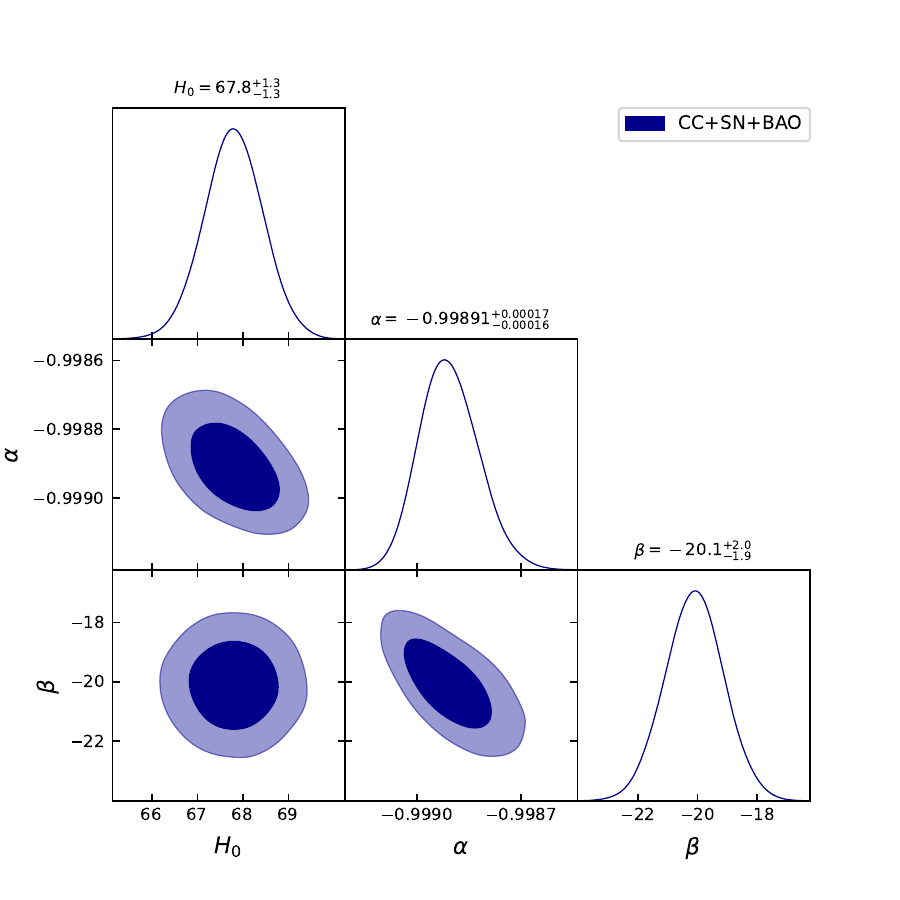}
\caption{The likelihood contours at $1-\sigma$ and $2-\sigma$ for the model parameters are obtained by combining the CC, BAO, and Pantheon datasets.}\label{C}
\end{figure}

\begin{figure}[H]
\centering\includegraphics[width=18cm,height=5.5cm]{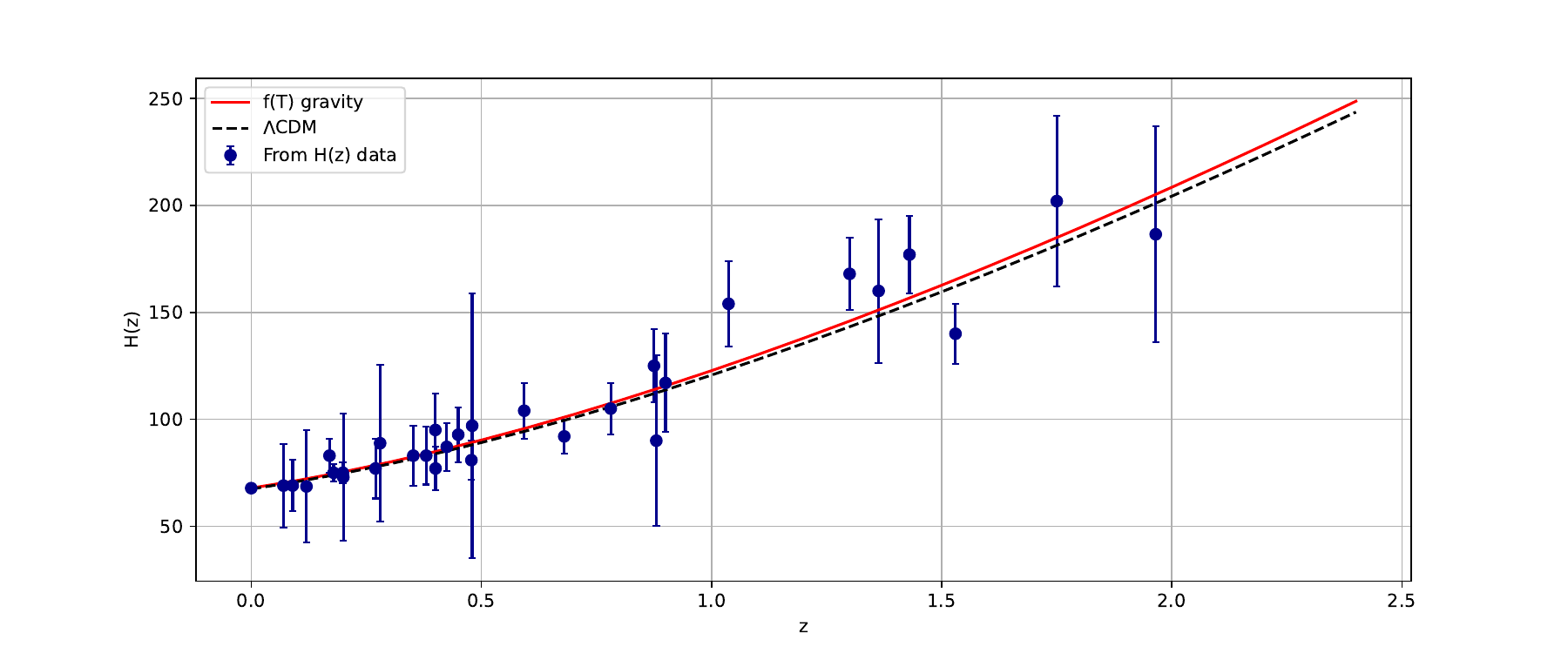}
\caption{The evolution of the Hubble parameter versus redshift $z$. The red line represents the model's curve, while the black dotted line corresponds to the $\Lambda$CDM model with $H_{0}=67.4 \pm 0.5$ and $\Omega_{m0}=0.315 \pm 0.007$ \cite{Planck/2018}. The blue dots with error bars represent the 31 sample points of $H(z)$.}\label{F_H}
\end{figure}

\end{widetext}

By minimizing the chi-square function for the combined $CC+SN+BAO$ datasets, we depict the likelihood contour at $1-\sigma$ and $2-\sigma$ confidence levels of the cosmological $f(T)$ model in Fig. \ref{C}. In summary, we estimate the best-fit value of the Hubble constant as $H_0=67.8^{+1.3}_{-1.3}$ km/s/Mpc, and the model parameters as $\alpha=-0.99891^{+0.00017}_{-0.00016}$ and $\beta=-20.1^{+2.0}_{-1.9}$. Recently, there have been reported anomalies in the measurement of the Hubble parameter between the Planck mission and independent cosmological probes. These discrepancies are commonly referred to as the Hubble constant tension ($H_0$ tension). In the Planck measurements \cite{Planck/2018}, the authors estimated $H_0 = 67.4 \pm 0.5 \, \text{km/s/Mpc}$, while in the study by Riess et al. \cite{Riess/2019}, the Hubble constant was found to be $H_0 = 74.03 \pm 1.42 \, \text{km/s/Mpc}$. It's worth noting that there is a significant $4.4\sigma$ tension in the measurement of the Hubble parameter between the Planck \cite{Planck/2018} and the findings of Riess et al. \cite{Riess/2019}. Comparing our results with those of the Planck measurements, we find that our estimated value of $H_0$, obtained from the combined $CC+SN+BAO$ datasets, exhibits a $0.29\sigma$ tension with the Planck results for $H_0$. Some important investigations on the Hubble constant tension are discussed in \cite{Valentino/2021A,Valentino/2021B,Yang/2021}.

\subsection{Cosmological parameters}

The evolution of various cosmological parameters for our cosmological $f(T)$ model, including the Hubble parameter, energy density parameter, deceleration parameter, and EoS parameter, is presented below based on the constrained values of the model parameters obtained from our analysis. 

The evolution of the Hubble parameter $H$ versus redshift $z$ is a fundamental aspect of cosmology, which sheds light on the universe's rate of expansion over time. In the context of our cosmological $f(T)$ model, the behavior of $H(z)$ can reveal important information about the underlying dynamics and properties of the universe. In Fig. \ref{F_H}, the Hubble parameter is shown to increase with cosmic redshift, indicating the universe's expanding nature. The increasing trend of $H(z)$ is likely due to DE, which is thought to drive the current accelerated expansion of the universe. In addition, we compared our model with the $\Lambda$CDM model by examining the evolution of the Hubble parameter $H(z)$ with the constraint values of our model parameters. Our $f(T)$ model aligns well with observational results and closely resembles the profile of the $\Lambda$CDM model.

\begin{figure}[H]
\includegraphics[scale=0.7]{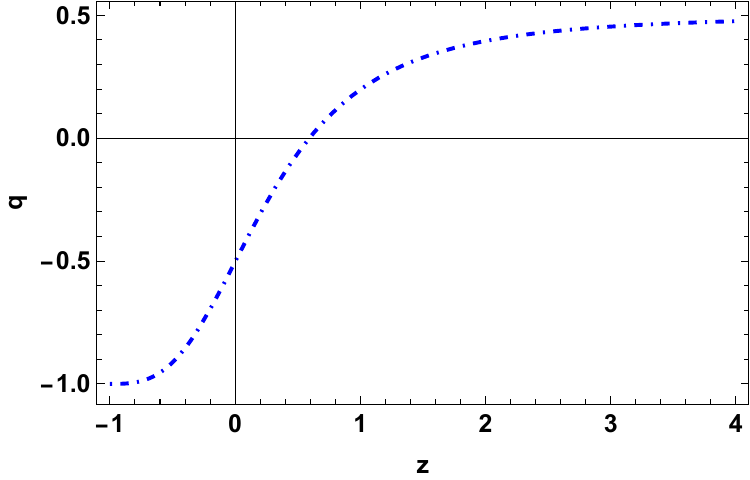}
\caption{The evolution of the deceleration parameter versus redshift $z$.}\label{F_q}
\end{figure}

Based on cosmological observations, the acceleration of the universe is a relatively recent development. To fully understand the universe's evolution, a cosmological model needs to encompass both its decelerating and accelerating phases. Therefore, studying the deceleration parameter $q$ is crucial for comprehending the universe's dynamics across its history. From Fig. \ref{F_q}, it is evident that the deceleration parameter is positive in the early stages of the universe and becomes negative in the late-time universe. This behavior suggests a transition from decelerated to accelerated expansion at a transition redshift value of $z_t=0.60$ \cite{Cunha1,Cunha2,Nair}. The deceleration parameter $q(z)$ increases as cosmic redshift progresses.

\begin{figure}[h]
\includegraphics[scale=0.7]{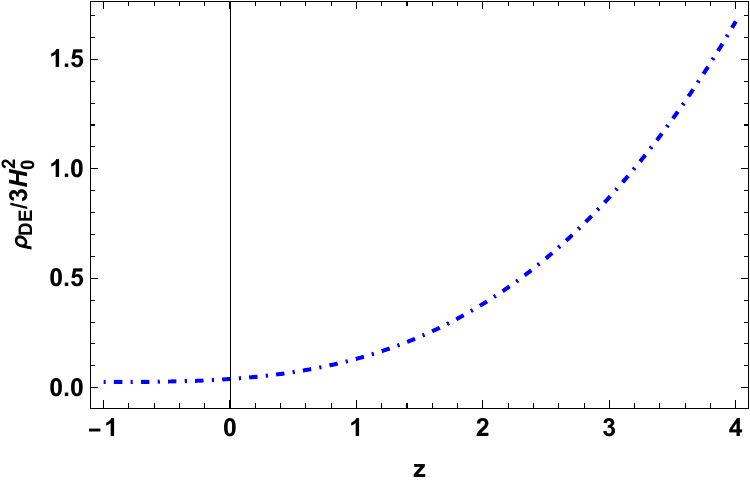}
\caption{The evolution of the density of the DE component versus redshift $z$.}\label{F_rho}
\end{figure}

From Fig. \ref{F_rho}, it is evident that the energy density of the DE component remains positive throughout the evolution of the universe and decreases with time, or equivalently, with decreasing redshift, eventually approaching zero in the distant future (i.e., as $z\rightarrow -1$). This behavior indicates that our model corresponds to a scenario where the DE density decays over time. 

\begin{figure}[h]
\includegraphics[scale=0.7]{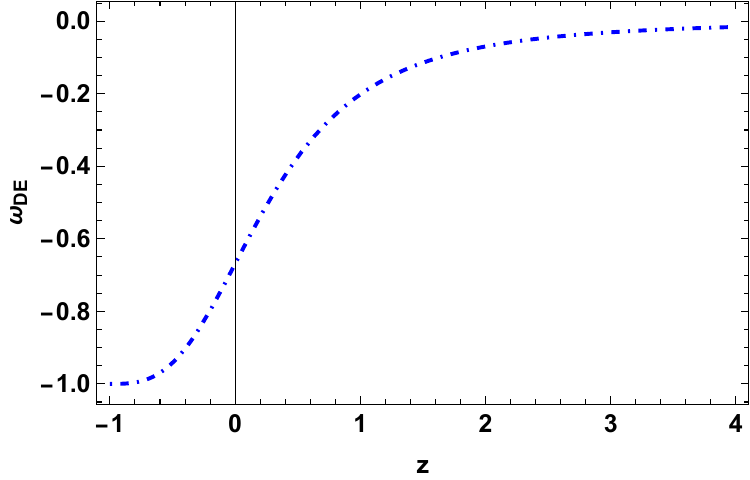}
\caption{The evolution of the EoS parameter of the DE component versus redshift $z$.}\label{F_wDE}
\end{figure}

\begin{figure}[h]
\includegraphics[scale=0.7]{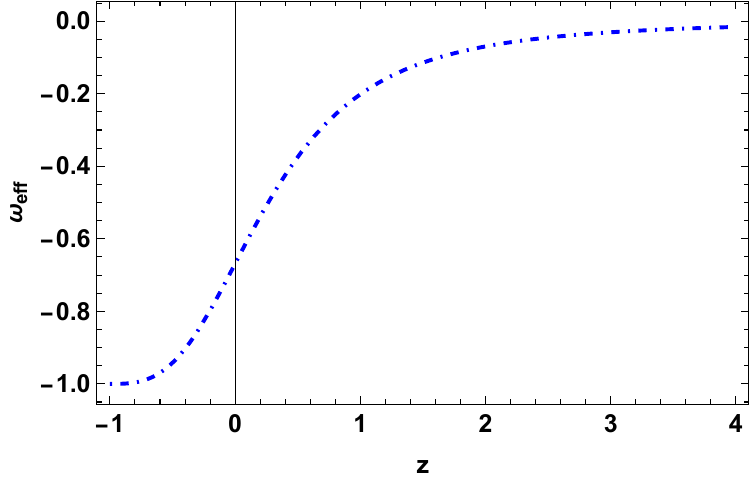}
\caption{The evolution of the effective EoS parameter versus redshift $z$.}\label{F_w}
\end{figure}

In addition, the EoS parameter is a crucial tool for categorizing the various epochs of accelerated and decelerated expansion of the universe. It helps identify different phases such as matter, radiation, quintessence, phantom, and the cosmological constant. For instance, a stiff fluid corresponds to $\omega = 1$. The radiation-dominated phase is represented by $\omega = \frac{1}{3}$. In contrast, the matter-dominated phase corresponds to $\omega = 0$. In quintessence, the EoS parameter is in the range $-1 < \omega < -\frac{1}{3}$, representing a form of DE where the energy density decreases as the universe expands. This phase typically leads to accelerated expansion. Phantom DE corresponds to an equation of state parameter $\omega < -1$. This scenario leads to even faster expansion and has exotic properties, such as negative kinetic energy for the scalar field. The cosmological constant, represented by $\omega = -1$, corresponds to a constant energy density that does not change as the universe expands. This leads to a scenario of constant acceleration, as observed in the standard $\Lambda$CDM model. From Figs. \ref{F_wDE} and \ref{F_w}, we note that the EoS parameter of the DE component, evolving due to torsion, and the effective EoS parameter exhibit behavior similar to quintessence. The current value of the EoS parameter for the DE component, corresponding to the constrained values of the model parameters, is $\omega_0 = -0.68$ \cite{O1,O2,O3}.

\section{Stability analysis using scalar perturbations}\label{sec5}

In this section, we investigate the stability analysis of the cosmological $f(T)$ model using the scalar perturbation approach. This is necessary due to the various assumptions made to derive the behavior of the Universe, making it challenging to assess their degree of generality. To validate the results, it is essential to examine the qualitative aspects of the field equations. We focus on linear, homogeneous, and isotropic perturbations to assess the stability of the cosmological solutions obtained. Here, we analyze the perturbations in the Hubble parameter and energy density using the perturbation geometry functions $\delta(t)$ and matter functions $\delta_{m}(t)$, following previous studies \cite{Wu12,Izumi13,Golovnev18}.
\begin{equation}\label{20}
H(t)\rightarrow H(t)(1+\delta),  \hspace{1cm}    \rho(t)\rightarrow \rho(t)(1+\delta_{m}).
\end{equation}

The linear perturbations of the function $f$ and its derivatives are given by $ \delta f = f_{T}\delta T$ and $\delta f_{T}= f_{TT}\delta T$, respectively. Here, the perturbation of the torsion scalar is represented as $T=-6H^{2}=-6H^{2}(1+\delta(t))^{2}=T(1+2\delta(t))$. By applying the perturbative approach to the equivalent FLRW equation in the background, we derive,
\begin{equation}\label{21}
-T(1+f_{T}-12H^{2}f_{TT})\delta=\kappa^{2}\rho \delta_{m}.
\end{equation}

The above equation illustrates the connection between matter and geometric perturbations, including the perturbed Hubble parameter. Nevertheless, Eq. \eqref{21} alone is inadequate for obtaining the analytical expression for the perturbation functions. Therefore, it is more appropriate to utilize the perturbation continuity equation, which can be expressed as
\begin{equation}\label{22}
\dot\delta_{m}(t)+3H(1+\omega_{eff})\delta(t)=0.
\end{equation}

This simplifies to:
\begin{equation}\label{23}
\delta(t)=\frac{1}{2T}. \frac{T+2Tf_{T}-f}{1+f_{T}+2Tf_{TT}} \delta_{m}(t).
\end{equation}

Then, the differential equation \eqref{22} is
\begin{equation}\label{24}
\dot\delta_{m}(t)+\frac{3H}{2T}(1+\omega_{eff})\frac{T+2Tf_{T}-f}{1+f_{T}+2Tf_{TT}} \delta_{m}(t)=0.
\end{equation}

Applying the method of separation of variables to the first-order ordinary differential equation in $\delta_{m}$, we obtain
\begin{equation}\label{25}
\delta_{m}(t)=\exp\left[-\frac{3}{2}(1+\omega_{eff})\int \frac{H}{T} \frac{T+2Tf_{T}-f}{1+f_{T}+2Tf_{TT}} dt\right].
\end{equation}

Using the zeroth-order $tt$-component of the Friedmann equations \eqref{F1} with \eqref{F2} and the continuity equation for a perfect fluid, we can obtain
\begin{equation}\label{26}
\dot{H} =\frac{T+2Tf_{T}-f}{4(1+f_{T}+2Tf_{TT})}(1+\omega_{eff}). 
\end{equation}

Now, by using Eqs. \eqref{25} and \eqref{26}, we can derive the expression for $\delta_m$ as
\begin{equation}\label{27}
\delta_{m}(t)=\exp\left[\int \frac{\dot{H}}{H}dt\right]=\exp \left[\int\frac{dH}{H} \right]=C_{1}H,
\end{equation}
where $C_{1}$ is an integration constant. At the present moment, $C_{1}=\frac{\delta_{m}(t_{0})}{H_{0}}$. Then, we can calculate $\delta(t)$ as
\begin{equation}\label{28}
\delta(t)=-\frac{C_{1}}{3(1+\omega_{eff})}\frac{\dot{H}}{H}.
\end{equation}

\begin{figure}[H]
\includegraphics[scale=0.7]{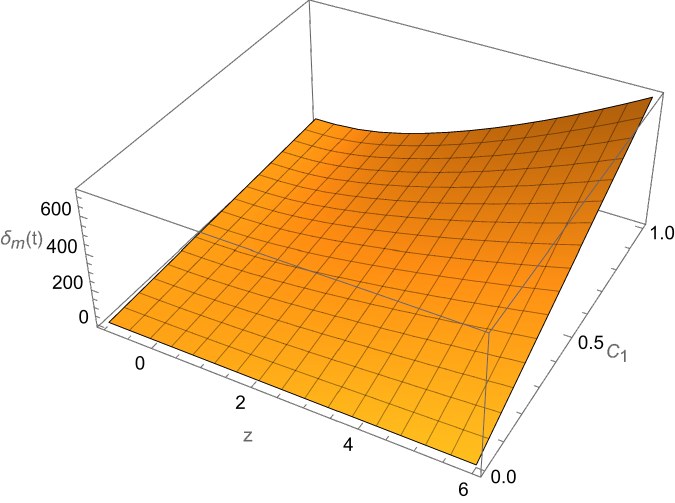}
\caption{The evolution of the matter perturbation versus redshift $z$ and $C_1$.}\label{F_dm}
\end{figure}

\begin{figure}[H]
\includegraphics[scale=0.7]{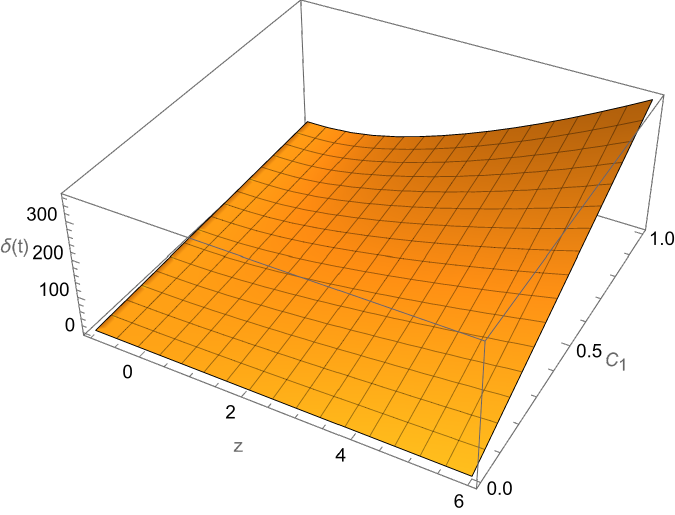}
\caption{The evolution of the Hubble perturbation versus redshift $z$ and $C_1$.}\label{F_d}
\end{figure}

This simplification shows that both $\delta_m(t)$ and $\delta(t)$ can be expressed in terms of cosmic redshift. In Fig. \ref{F_dm}, we depict the behavior of the perturbed energy density, and in Fig. \ref{F_d}, we depict the behavior of the perturbed Hubble parameter. Both $\delta(t)$ and $\delta_m(t)$ exhibit a decaying trend over time, approaching zero at late times (i.e., as $z\rightarrow -1$). This indicates that the cosmological $f(T)$ model is stable under the scalar perturbation approach \cite{Duchaniya/2022,Mussatayeva/2023}.

\section{Conclusion} \label{sec6}

Comprehending the DE evolution is a major difficulty in modern cosmology. The nature of DE, responsible for the accelerated expansion of the universe, remains one of the most profound mysteries in physics. Various theoretical models, including the $\Lambda$CDM model and alternative theories such as $f(T)$ gravity, have been proposed to explain DE. These models aim to describe the properties and evolution of DE, which is crucial for understanding the past, present, and future of our universe.

In this study, we have focused on a specific $f(T)$ cosmological model and analyzed its behavior using observational data, including 31 data points from the CC dataset, 1048 points from the Pantheon SNe samples, and 6 points from the BAO dataset. Specifically, we considered a linear $f(T)$ model with an additional constant term, i.e. $f(T)=\alpha T+\beta$, where $\alpha$ and $\beta$ are parameters of the model. Then, we derived the expression for the Hubble parameter as a function of cosmic redshift for non-relativistic pressureless matter. By minimizing the chi-square function for the combined $CC+SN+BAO$ datasets, we obtained the best-fit values for the Hubble constant, $H_0$, and the model parameters $\alpha$ and $\beta$. Our analysis yielded $H_0=67.8^{+1.3}_{-1.3}$ km/s/Mpc, $\alpha=-0.99891^{+0.00017}_{-0.00016}$, and $\beta=-20.1^{+2.0}_{-1.9}$. Kumar et al. \cite{Kumar/2023} presented new cosmological constraints on $f(T)$ gravity based on full Planck-CMB and SNe Ia data. The authors found $H_0$ to be $66.2^{+1.2}_{-1.4}$ for BAO+BBN and $66.8 \pm 1.2$ for BAO+BBN+Pantheon, which is consistent with our values. The evolution of the Hubble parameter, shown in Fig. \ref{F_H}, indicates an increasing trend with cosmic redshift, consistent with an expanding universe. The deceleration parameter, illustrated in Fig. \ref{F_q}, transitions from positive to negative values, indicating a shift from decelerated to accelerated expansion at a transition redshift of $z_t=0.60$.

Furthermore, the energy density of the DE component, as depicted in Fig. \ref{F_rho}, decreases over time and approaches zero in the far future, suggesting a decaying DE model. The EoS parameter for the DE component, shown in Figs. \ref{F_wDE} and \ref{F_w}, exhibits behavior similar to quintessence, with $\omega_0=-0.68$. Finally, our stability analysis using the scalar perturbation approach indicates that the cosmological $f(T)$ model is stable, with both $\delta(t)$ and $\delta_m(t)$ decaying over time and approaching zero at late times. The obtained results therefore show that our cosmological $f(T)$ model effectively generates the necessary conditions to explain late-time cosmic acceleration without requiring the presence of a DE component in the matter composition. In conclusion, the study of DE is a complex and challenging field that requires the combined efforts of theorists and observers. By continuing to study the evolution of DE using innovative theoretical models and precise observational data, we can hope to unravel the mysteries of the universe and gain a deeper understanding of its underlying principles.

\section*{Data Availability Statement}
There are no new data associated with this article.

\end{document}